\documentclass[prb,twocolumn,superscriptaddress,floatfix,showpacs,amsmath,amssymb]{revtex4-2}

\usepackage{graphicx}
\usepackage{dcolumn,xcolor}
\usepackage{bm}
\usepackage{color}
\usepackage{tabularx}
\usepackage{siunitx}
\usepackage{verbatim}
\usepackage{parskip}
\usepackage{soul,xr}
\usepackage[normalem]{ulem}
\newcommand{\gio}{GdInO$_3$}

\newcommand{\rmo}{$R$MnO$_3$}
\newcommand{\rio}{$R$InO$_3$}
\newcommand{\tio}{TbInO$_3$}
\newcommand{\tn}{$T_{\rm N}$}
\newcommand{\tntwo}{$T_{\rm N2}$}
\newcommand{\bpab}{$B\parallel {\rm ab}$}
\newcommand{\bpc}{$B\parallel {\rm c}$}
\newcommand{\mb}{$\mu_{\rm B}$}
\newcommand{\mbfu}{\(\mu _{\rm B}/{\rm f.u.}\)}

\newcommand{\jmk}{J/(mol\,K)}
\newcommand{\E}{0.22~erg/(G$^2$mol)}

\newcommand{\rp}{$R_{\rm p}$}
\newcommand{\rwp}{$R_{\rm wp}$}
\newcommand{\bcone}{$B_{\rm c1}$}
\newcommand{\bconeupper}{$B_{\rm c1}^{\triangledown}$}

\newcommand{\bconelower}{$B_{\rm c1}^{\vartriangle}$}
\newcommand{\baone}{$B_{\rm a1}$}
\newcommand{\batwo}{$B_{\rm a2}$}
\newcommand{\bathree}{$B_{\rm a3}$}
\newcommand{\bs}{$B_{\rm s}$}
\newcommand{\bctwo}{$B_{\rm c2}$}
\newcommand{\tm}{$T^*$}
\newcommand{\Cp}{$c_{p}$}

\begin{document}

\preprint{APS/123-QED}

\title{1/3 plateau and 3/5 discontinuity in the magnetization and the magnetic phase diagram of hexagonal \gio\ }

\author{N.~Yuan}
\email{ning.yuan@kip.uni-heidelberg.de}
\affiliation{Kirchhoff Institute of Physics, Heidelberg University, INF 227, D-69120 Heidelberg, Germany}
\author{A.~Elghandour}
\affiliation{Kirchhoff Institute of Physics, Heidelberg University, INF 227, D-69120 Heidelberg, Germany}
\author{W.~Hergett}
\affiliation{Kirchhoff Institute of Physics, Heidelberg University, INF 227, D-69120 Heidelberg, Germany}
\author{R.~Ohlendorf}
\affiliation{Kirchhoff Institute of Physics, Heidelberg University, INF 227, D-69120 Heidelberg, Germany}
\author{L.~Gries}
\affiliation{Kirchhoff Institute of Physics, Heidelberg University, INF 227, D-69120 Heidelberg, Germany}
\author{R.~Klingeler}
\email{klingeler@kip.uni-heidelberg.de}
\affiliation{Kirchhoff Institute of Physics, Heidelberg University, INF 227, D-69120 Heidelberg, Germany}

\date{\today}

\begin{abstract}

We report the high-pressure optical floating-zone growth of \gio\ single crystals and show its magnetic phase diagram down to the mK-regime as determined by magnetization measurements. The centered-honeycomb lattice structure shows considerable magnetic frustration ($\lvert\Theta\rvert /T_{\rm N}\simeq 5$) and develops long-range magnetic order below \tn~=~2.1~K from a short-range ordered paramagnetic phase. Concomitantly, a small net magnetic moment evolves at \tn\ which points along the crystallographic $c$ direction. Upon cooling, the net moment reorients at $T^{**}\simeq 1.7$~K and $T^{*}\simeq 1$~K. A broad 1/3 plateau indicative of the up-up-down ($uud$) spin configuration appears for $B||c$ but is absent for $B||ab$, thereby suggesting easy axis anisotropy. At $T=0.4$~K, a jump in magnetization at $\simeq 3/5$ of the saturation magnetization signals a discontinuous transition to a high field phase and we find evidence for a possible tricritical point. Small energy and field scales in the accessible regimes render \gio\ a prime example to study the phase diagram of a semiclassical frustrated hexagonal lattice in the presence of weak easy axis anisotropy of mainly dipolar origin.

\end{abstract}

\maketitle

\section{Introduction}
Due to a wealth of relevant physical properties, perovskite rare earth (RE) oxides with the nominal composition $R$BO$_3$ ($R$ a RE ion) form a versatile class of materials and are a major topical research area in condensed matter physics and materials science. Prominent examples of spectacular phenomena observed in this class of materials are metal-insulator transitions in rare earth nickelates~\cite{REN-mercy2017structurally,REN-sanchez1996metal}, colossal magnetoresistance, phase separation, or charge and orbital order in manganates~\cite{LSMO-ramirez1997colossal,uehara1999percolative,Rao1998,uhlenbruck1999interplay} or multiferroicity in $R$MnO$_3$~\cite{YMO-choi2010insulating,YMO-van2004origin}. Similar as hexagonal \rmo , the hexagonal \rio\ ($R$ = Eu, Gd, Tb Dy and Ho) systems crystallize in the $P6_3cm$ space group~\cite{RIO2018nonequivalent, RIO2020first}. Their centered honeycomb lattice structure renders \rio\ an excellent platform to study geometrically frustrated magnets. 
Prior to 2017, studies on \rio\ were limited to polycrystalline samples~\cite{GIO2017geometrically}. Accessibility of macroscopic single crystals such as \gio~\cite{GIO2018laser}, \tio~\cite{TIO2019NP,TIO2019PRX}, and Mn-doped \tio~\cite{TIMO2019PRB} has boosted the field as for example ferroelectricity and spin liquid behavior was found in \tio~\cite{TIO2019NP,TIO2019PRX}. In comparison to the quantum spin liquid candidate \tio , the properties of \gio\ resemble more that of as classical magnet~\cite{GIO2021magnetic}. The appearance of a weak feature in the isothermal magnetization curves was interpreted as an indication of a  1/3 magnetization plateau which is a typical feature of the up-up-down ($uud$) phase in triangular antiferromagnets~\cite{seabra2011,BMNO2014plateau,BCSO2013plateau, RFMO2006plateau}. In addition, \gio\ features ferroelectricity as confirmed by observation of the $P$($E$) hysteresis loop as well as a $Z_6$ vortex topological domain structure~\cite{GIO2018laser}.

Volatilization of In$_2$O$_3$ has long been a major challenge for the preparation of \gio\ single crystals. In this work, we have mitigated this issue by employing the high-pressure optical floating-zone method and show that high-quality \gio\ single crystals are successfully grown when using a high oxygen pressure of 30~bar. Using the single crystals we have constructed the magnetic phase diagrams in the temperature regime down to 400~mK and in magnetic fields up to 14~T. In zero magnetic field, distinct anomalies in the magnetization and specific heat signal the evolution of long-range magnetic order at \tn\ = 2.1~K. Applying magnetic fields $B||c$ axis yields a 1/3 magnetization plateau in the isothermal magnetization which is centered at about 3~T. This magnetization plateau behavior is absent for $B||ab$ plane. The system also exhibit a small net magnetic moment along $c$ axis, weak easy-axis anisotropy, re-orientation processes both in zero magnetic field and driven by the field, and a discontinuous transition into a high-field phase.

\section{Experimental Methods}

Polycrystalline \gio\ was synthesized by a standard solid-state reaction following Refs.~\cite{GIO2018laser,GIO2021magnetic}. Stoichiometric amounts of Gd\(_2\)O\(_3\) and In\(_2\)O\(_3\) powders were well mixed and calcined at 1350~$^\circ$C for 24~h (air flow, ambient pressure). The resulting material was grinded and sintered for three times to ensure complete reaction. Polycrystalline rods were prepared by hydrostatically pressing the powders under the pressure of 60~MPa and annealing them for 36~h at 1400~$^\circ$C. \gio\ single crystal was grown by using the high-pressure optical floating-zone furnace (HKZ, SciDre) as described below. The phase purity and crystallinity were studied by powder X-ray diffraction (XRD) and the back-reflection Laue method. XRD was performed at room temperature by means of a Bruker D8 Advance ECO diffractometer using Cu-K$\alpha$ radiation ($\lambda$ = 1.5418~\AA). Data have been collected in the 2$\Theta$ range of 10 – 90$^\circ$ with 0.02$^\circ$ step-size. Structure refinement was carried out using the FullProf Suite by means of the Rietveld method \cite{rodriguez2001introduction}. Magnetic studies at 1.8 - 300~K have been performed in a SQUID magnetometer (MPMS3, Quantum Design Inc.) and employing the vibrating sample magnetometer option of a Physical Properties Measurement System (PPMS, Quantum Design Inc.). For studies at temperatures between 0.4 and 5~K, the iQuantum $^{3}$He option of MPMS3 was used. A relaxation method was used to perform specific heat measurements in the PPMS.

\section{Results}

\subsection{\gio\ single crystal growth}

Heavy volatilization of In$_2$O$_3$ and low surface tension of the melts challenges growth of macroscopic \gio\ single crystals. To suppress volatilization, the crystals reported here were grown under an oxygen pressure of 30~bar using the high-pressure floating-zone furnace (HKZ, SciDre)~\cite{Wizent2009,Neef2017}. High pressure was maintained at an O$_2$ flow rate of 0.1~l/min. A Xenon arc lamp operating at 5~kW was employed and the growth was performed inside a sapphire chamber. A relatively fast growth rate of 10~mm/h was chosen in order to further mitigate In\(_2\)O\(_3\) volatilization. At slower growth rates, we observed significant amounts of deposited In$_2$O$_3$ volatiles (see Fig.~S1a of the Supplemental Material~\cite{SM}) adhering to the inner protection glass tube, thereby affecting the focusing of light and preventing stable growth.

\begin{table}[htb]
\caption{\label{tab:table1}
Growth parameters and phase analysis from the Rietveld refinement of the room temperature powder XRD data of GdInO\(_3\) single crystals from the literature~\cite{GIO2018laser,GIO2021magnetic} and reported at hand (HKZ).
}

\begin{ruledtabular}
\begin{tabular}{cccccccc}
& Laser\footnotemark[1] & Two-mirror\footnotemark[2] & HKZ\footnotemark[3] \\
\hline
Atmosphere& O\(_2\) & O\(_2\) & O\(_2\) \\
Flow rate(l/min)& 0.1 & 0.2 & 0.1 & \\
O\(_2\) Pressure (bar)& 9.5 & 9 & 30 & \\
Growth rate (mm/h)& 5-10 & 10 & 10 & \\
Lattice parameter \emph a(\AA)& 6.3301(4) & 6.3433(3) & 6.3451(3)  \\
Lattice parameter \emph c(\AA)& 12.3340(1) & 12.3320(1) & 12.3408(9) \\
\end{tabular}

\end{ruledtabular}
\footnotetext[1]{Grown by the laser floating zone furnace (Crystal Systems Inc.), see Ref.~\cite{GIO2018laser}.}
\footnotetext[2]{Grown by a two-mirror optical floating zone furnace (IRF01-001-05, Quantum Design), see Ref.~\cite{GIO2021magnetic}.}
\footnotetext[3]{Grown by the high-pressure optical floating zone furnace (HKZ, SCIDRE), this work.}
\end{table}

Using an in-situ temperature measurement by means of a
two-color pyrometer~\cite{Hergett2019,Hergett2021}, the temperature of the melting zone during the growth was determined to about 1750~$^\circ$C. The feed and seed rods were counter-rotated at 20~rpm to improve homogeneity of the melt; both feed and seed rods were pulled at 10~mm/h. The obtained boule is shown in Fig.~S1b in the SM~\cite{SM}. Table~\ref{tab:table1} lists the growth parameters used in this work and those in previous studies, as well as the refined lattice parameters and further characteristics of the respective crystals.

\begin{figure}[htb]
\centering
\includegraphics [width=\columnwidth,clip] {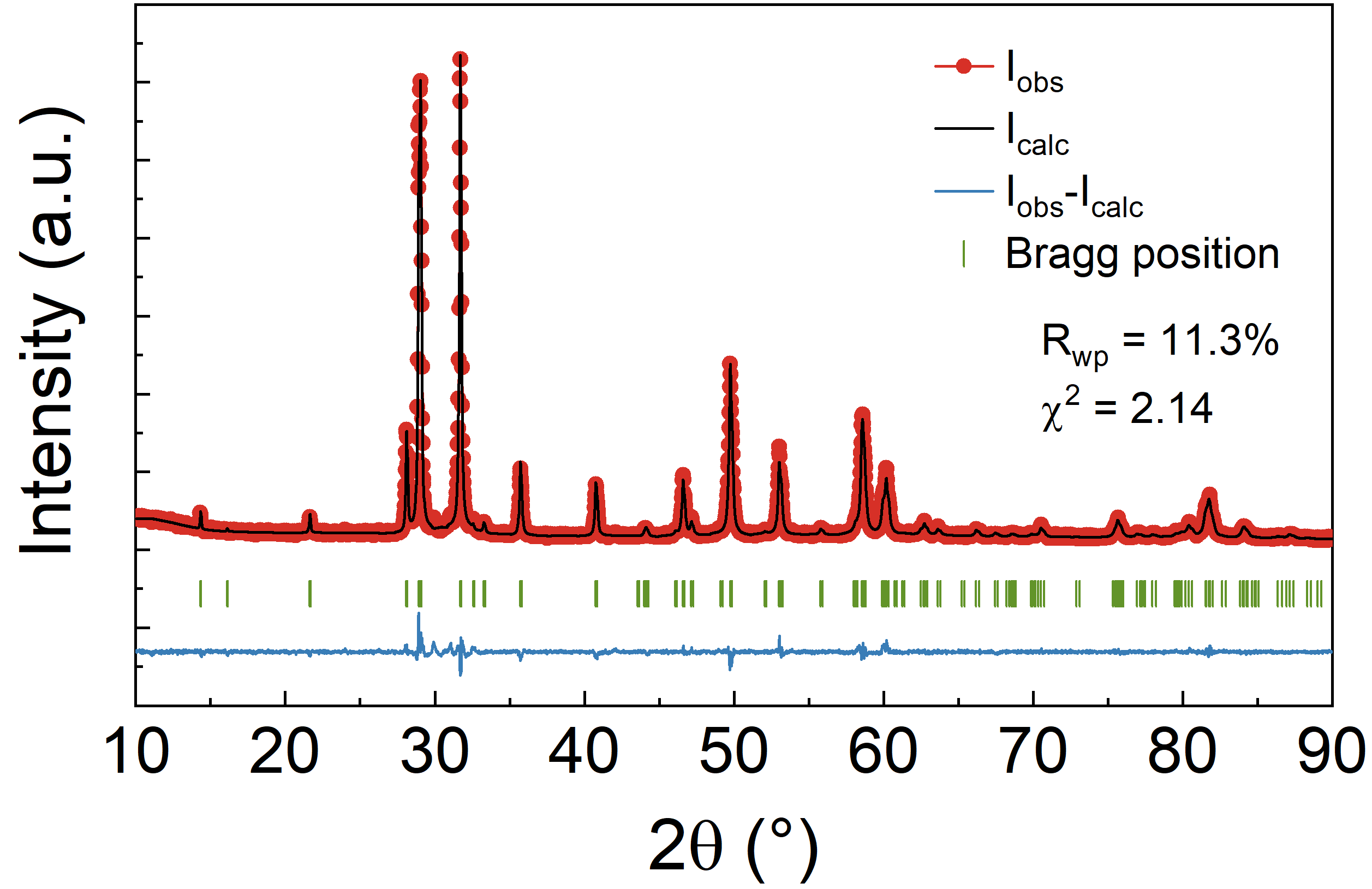}
\caption{Room temperature XRD pattern and corresponding Rietveld refinement of powdered \gio\ single crystals grown at 30~bar O$_2$ pressure. The observed diffraction pattern is shown in red, the calculated one in black, and the difference between them is shown in blue. Refinement is based on the hexagonal crystal system (space group $P6_3cm$, no. 185) of \gio\ as a main phase. The vertical green bars show the expected Bragg positions. The refinement converged to \rp = 12.0 $\%$, \rwp = 11.3 $\%$, $\chi^{2}$ = 2.14.} \label{fig.2}
\end{figure}

A powder x-ray diffractogram on a ground single crystal as well as an Rietveld refinement to the data is shown in Fig.~\ref{fig.2}. The result of the XRD refinement demonstrates that our sample is free of impurities, and the lattice parameters and the crystal structure match the reported crystals~\cite{GIO2021magnetic,GIO2018laser}. X-ray Laue diffraction in back scattering geometry was used to confirm single crystallinity and to orient the single crystals which
were then cut with respect to the crystallographic main
directions using a diamond-wire saw. Fig.~S1c~\cite{SM} shows the single crystal sample used for the magnetic and specific heat measurements. The Laue pattern in Fig.~S1d~\cite{SM} illustrates the high crystallinity of this sample. Laue diffraction performed at several other pieces of the \gio\ boule which were cleaved at room-temperature confirm that the dominant growth direction is in the $ab$ plane.

\begin{figure}[htb]
\centering
\includegraphics [width=\columnwidth,clip] {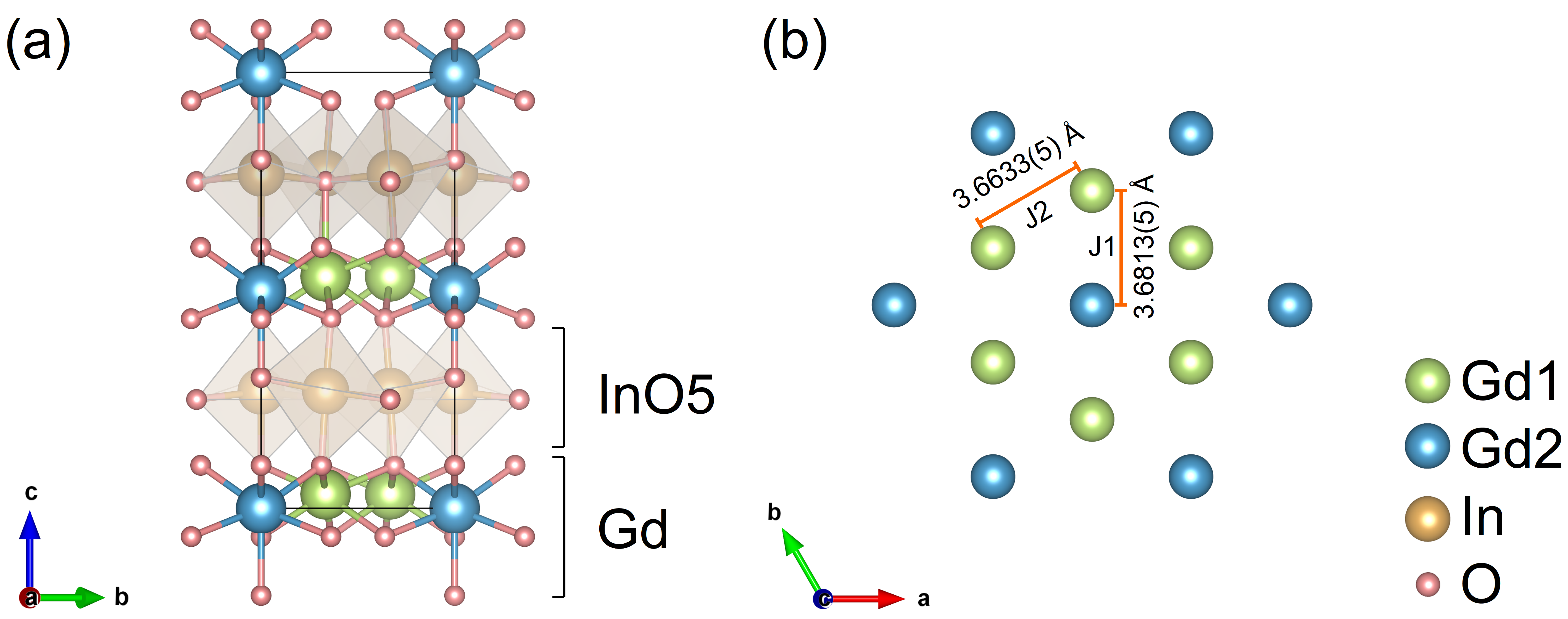}
\caption{(a) Schematic of the crystal structure of \gio\ along the [100] projection. (b) In-plane honeycomb arrangement of two inequivalent atomic sites of Gd$^{3+}$ ions. $J_1$ and $J_2$ represent two magnetic exchange interactions that are distinguished by symmetry. (COD No.7237332~\cite{GIO2018laser,momma2011vesta})} \label{fig.3}
\end{figure}

The crystal structure of \gio\ is shown in Fig.~\ref{fig.3}. It belongs to the hexagonal space group $P6_3cm$ (No.185), which alternately arranges corner connected layered InO\(_5\) bipyramids and Gd layers. The structure features two inequivalent Gd sites in the Gd layers with Wyckoff positions 2a and 4b, respectively. The two types of Gd sites form an arc-like arrangement when viewed from the [100] direction (see Fig.~\ref{fig.3}a). Hence, in the centered honeycomb layers formed by the Gd atoms in the plane perpendicular to the [001] axis, there are two slightly different Gd-Gd distances which may result in two distinct nearest-neighbor magnetic exchanges parameters, $J$\(_1\) (Gd1-Gd2) and $J$\(_2\) (Gd1-Gd1) (see Fig.~\ref{fig.3}b)~ \cite{GIO2017geometrically,GIO2021magnetic}.

\subsection{Magnetization $M(T,B)$}

\begin{figure}[htb]
\centering
\includegraphics [width=0.95\columnwidth,clip] {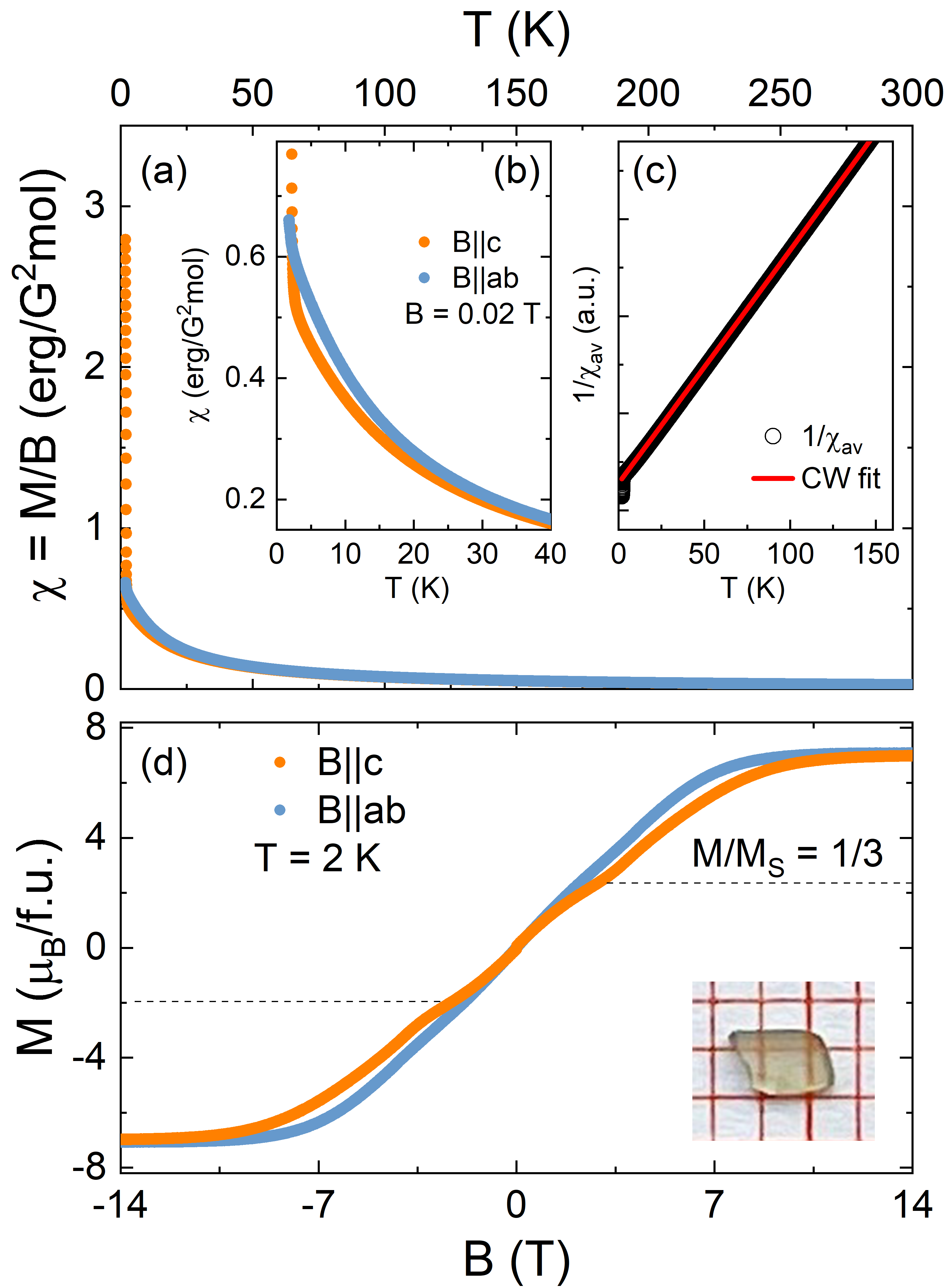}
\caption{(a) Temperature dependence of the static magnetic susceptibility $\chi = M/B$, obtained at $B=0.02$~T applied along the crystallographic $c$ axis (\bpc) and in-plane (\bpab), respectively; the inset (b) highlights small anisotropy at low temperatures. (c) Inverse of the averaged susceptibility $\chi_{\rm av}=(\chi_{c}+2\chi_{ab}$)/3 and Curie-Weiss fit (red line). (d) Isothermal magnetization at $T = 2$~K for \bpc\ and \bpab. The horizontal dashed line marks 1/3 of the saturation magnetization. The inserted picture shows the oriented single crystal under study on a mm-grid.} \label{Mintro}
\end{figure}

The static magnetic susceptibility $\chi = M/B$ obeys Curie-Weiss-like behaviour down to about 50~K as shown in Fig.~\ref{Mintro}a and c. At 300~K, $\chi(B||ab)/\chi(B||c) \simeq 1.007$ signals purely paramagnetic behaviour and negligible anisotropy of the $g$-factor. Upon cooling below $\sim 50$~K, a small anisotropy between $\chi_{\rm c}$ and $\chi_{\rm ab}$ evolves before an anomaly at \tn~=~2.1~K indicates the onset of long-range magnetic order (see Fig.~\ref{Mintro}b). Fitting the averaged susceptibility well above \tn\ by an extended Curie-Weiss law, i.e., $(\chi_{c}+2\chi_{ab})/3 = \chi_0 +  N_{\rm A}p_{\rm eff}^{2}/[3 k_{\rm B}(T-\Theta )]$ with the Avogadro number $N_{\rm A}$ and the Boltzman constant $k_{\rm B}$ yields an excellent agreement with the data (Fig.~\ref{Mintro}c). The fit yields the Weiss temperature $\Theta = -12(1)$~K and the effective magnetic moment $p_{\rm eff}= 8.2(1)$~\mb . This value is slightly larger than the theoretical value of 7.94~\mb\ for a free Gd$^{3+}$ moment. The obtained negative Weiss temperature $\Theta$ implies predominant antiferromagnetic interactions. We note the frustration parameter $f=\lvert\Theta\rvert /T_{\rm N}\simeq 5$ which suggests considerable spin frustration in \gio.

\begin{figure}[htb]
\centering
\includegraphics [width=\columnwidth,clip] {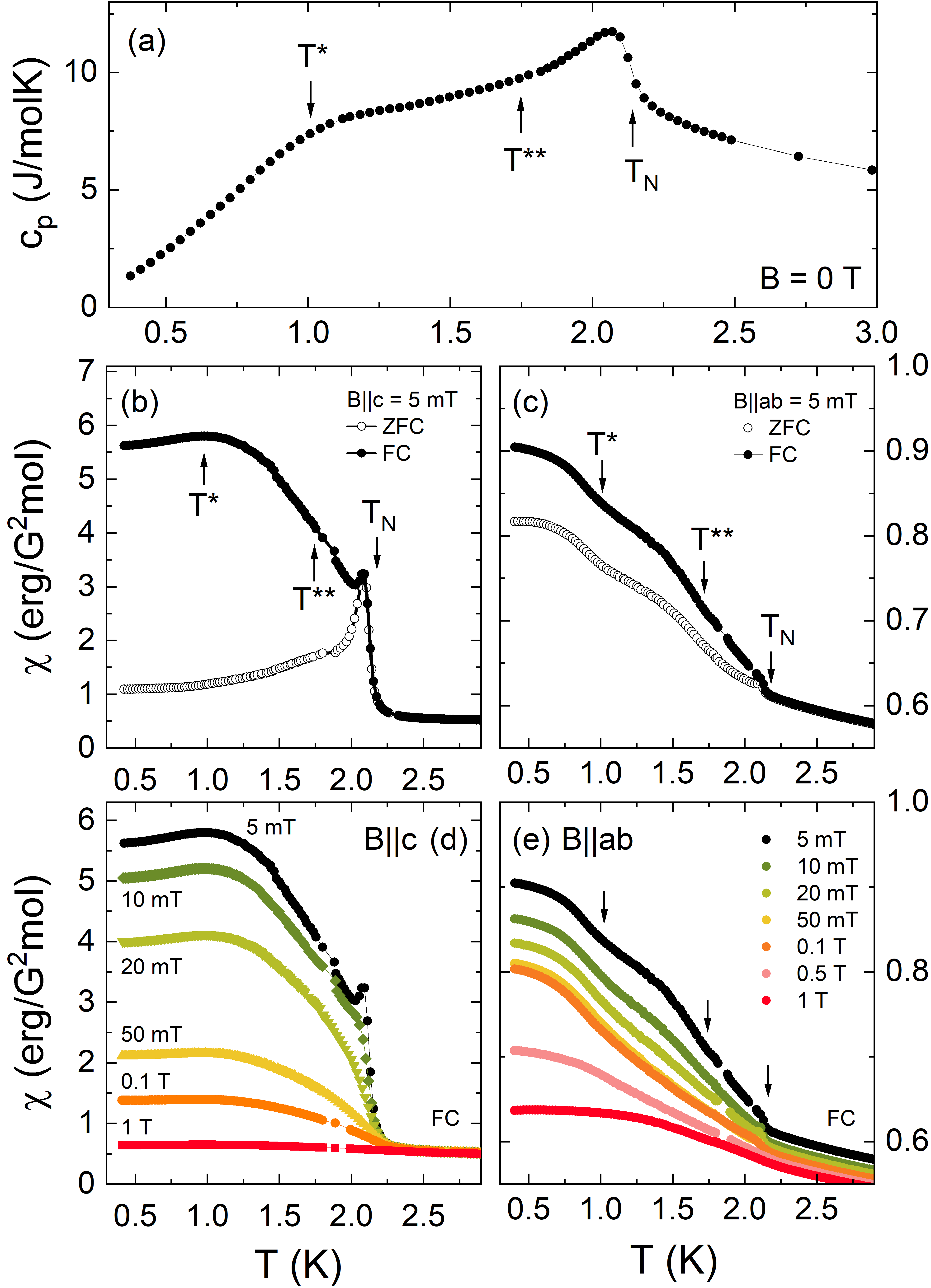}
\caption{(a) Specific heat at $B=0$~T and static magnetic susceptibility $\chi=M/B$ (b) in $B||c = 5$~mT and (c) $B||ab=5$~mT obtained in the (Zero-)Field-Cooled ((Z)FC) regime (black). (d-e) Static magnetic susceptibility (FC) at different external magnetic fields up to 1~T. \tn , $T^*$, and $T^{**}$ have been determined as described in the text.}
\label{chicp}
\end{figure}

The onset of long-range antiferromagnetic (AF) order is associated with a clear anomaly in the specific heat shown in Fig.~\ref{chicp}a. Concomitantly and in agreement to the presence of a hysteresis in $\chi$ vs. $T$, there is a steep increase of $\chi(B||c)$ at \tn\ and pronounced hysteresis between the data obtained after cooling in magnetic field (field-cooled; FC) and in zero magnetic field (zero-field-cooled; ZFC) (see Fig.~\ref{chicp}b,c). Upon cooling, the FC static susceptibility ($\chi_{\rm FC}$) increases further until a broad maximum develops around $T^{\rm *}\simeq 1$~K; below \tm , the magnetization $\chi(B||c)$ slightly decreases. Concomitantly, there is a broad hump in the specific heat which has been interpreted as signature of a second antiferromagnetic phase  appearing at \tntwo , with \tntwo = 1.05~K as indicated by the associated maximum in \Cp /$T$~\cite{GIO2021magnetic}. This hump is clearly visible in our data in Fig.~\ref{chicp}a, too. We note, however, that the hump neither signals an entropy jump nor is it $\lambda$-shaped and hence there is no clear signature of a thermodynamic phase transition at \tm . If one would alternatively interpret the hump as a broad jump in \Cp , an entropy-conserving construction would suggest \tntwo~$\simeq 1.4$~K. At this temperature, however, there is no anomaly in the magnetization. As will be discussed below, the observation of an anomaly in $\chi$ vs. $T$ and a maximum of a broad hump in \Cp\ at the same temperature \tm\ may hence be considered a signature of crossover associated with re-orientation of the spin structure. A re-orientation process is not necessarily associated with a proper thermodynamic phase boundary (see, e.g., rotation of a small ferromagnetic moment in Eu$_2$CuSi$_3$~\cite{cao2010magnetic} where a similar hump in \Cp\ is observed).

The magnetization data imply hysteresis between FC and ZFC measurements at low temperatures with bifurcation below $T_{\rm N}$. Both the increase of magnetization and hysteresis further confirm the presence of a weak ferromagnetic component below \tn . The tiny peak in $\chi_{\rm FC}$ (\bpc) is typical of a ferromagnetic-like domain state signaling decrease of magnetic anisotropy when heating towards the transition temperature. This scenario is supported by the observations in larger fields which suppress all features mentioned above, i.e., jump in magnetization, bifurcation, tiny peak, and the hump at \tm (see Fig.~\ref{chicp}b).

Several features appear in $\chi(B||ab)$ as demonstrated in Fig.~\ref{chicp}c: There is a kink at \tn\ indicating a very small increase of magnetization in the ordered phase. At \tm , there is a change in the slope as indicated by an inflection point in $\chi$ vs. $T$. In addition, we observe an anomaly at $T^{\rm **}\simeq 1.7$~K (see Fig.~\ref{chicp}c). The observation that $\chi(B||c)$ decreases at \tm\ while $\chi(B||ab)$ increases suggests the partial rotation of ferromagnetic component towards the $ab$ plane. The fact that \tm\ is also characterized by a broad hump in the specific heat (\ref{chicp}a) indicates that the changes in the magnetization in the ordered phase are associated with anomalous entropy changes. 
Our data do not allow us to trace the field dependence $T^{\rm **}($\bpab$)$. In contrast, $T^{\rm *}$ can be detected and it does not visibly change for small fields $B\leq 0.1$~T and increases for higher fields.

Further information on the long-range ordered phase is obtained by estimating the actual jump size of the specific heat anomaly at \tn . It is derived from the data by an entropy-conserving method to $\Delta c_{\rm p} = 4.1(3)$~\jmk~\cite{Gries2022}. This value is much smaller than the expected mean field value for a $S=7/2$ system of $\Delta c_{\rm p} = R\frac{5S(S+1)}{S^2+(S+1)^2} \simeq 20.1$~\jmk , with $R$ being the gas constant~\cite{morrish2001physical}. Note, that the actual jump
size associated with the measured anomaly can be even
smaller as it may be superimposed by critical fluctuations. The rather small specific heat anomaly observed experimentally implies significant short range magnetic order above \tn\ (as, e.g., suggested by the frustration parameter $f\simeq 5$) and/or considerable spin disorder below \tn .

\begin{figure}[hbt]
\centering
\includegraphics [width=\columnwidth,clip] {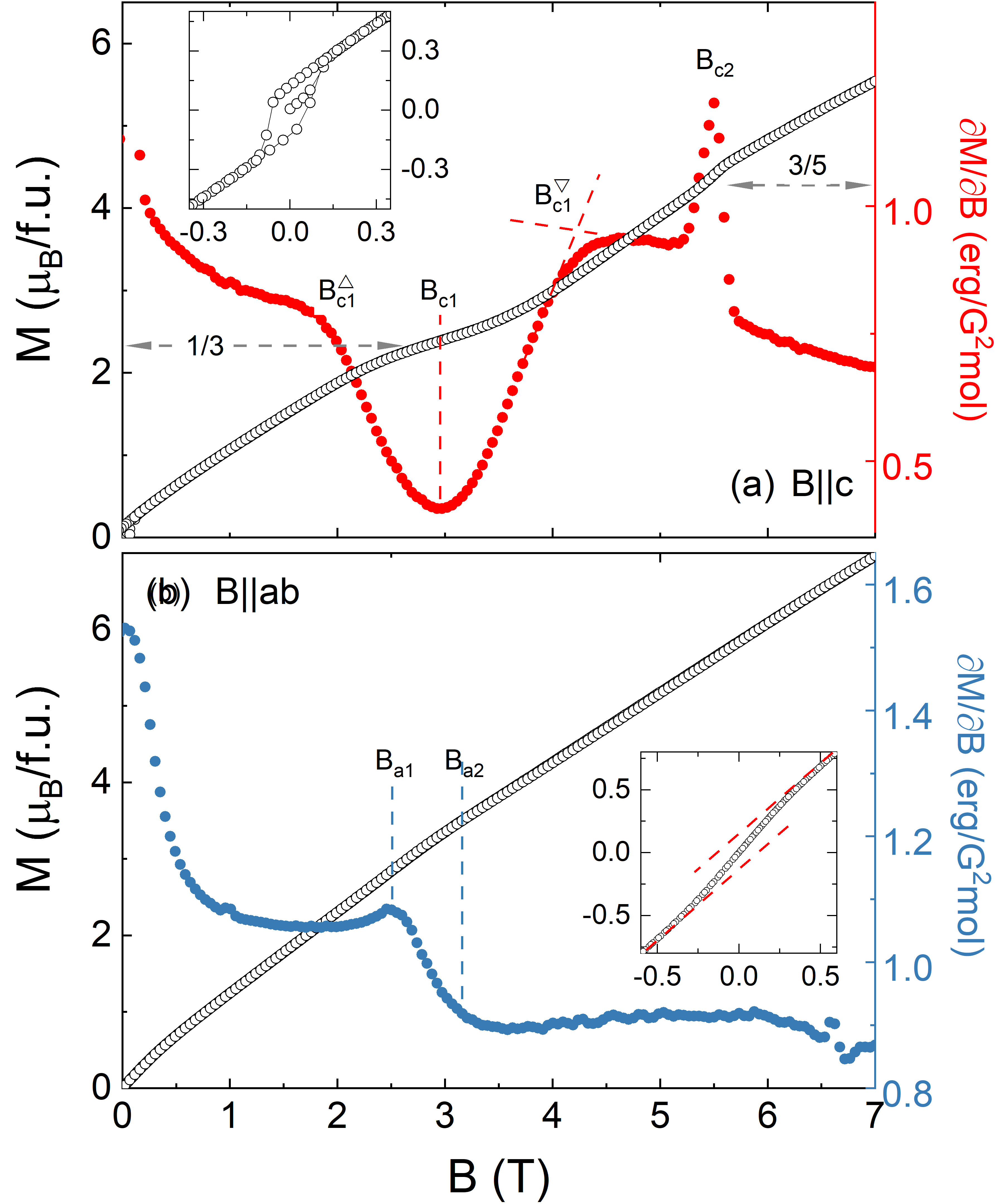}
\caption{Isothermal magnetization at $T = 0.4$~K for (a) $B||c$ axis and (b) $B||ab$ plane (left ordinates) and corresponding magnetic susceptibilities $\partial M/\partial B$ are shown (right ordinates). Insets highlight the behaviour around zero field. Horizontal dashed lines mark $M_s$/3 and $3M_s$/5; vertical lines show the anomaly fields as described in the text.} \label{M400mK}
\end{figure}

The effect of magnetic fields on the magnetic ground state is further illustrated by the isothermal magnetization $M$ vs. $B$ and the associated magnetic susceptibility $\partial M/\partial B$ at $T=2$~K (up to 14~T: Fig.~\ref{Mintro}d) and at $T=0.4$~K (up to 7~T: Fig.~\ref{M400mK}). From $M (T=2~\rm{K})$, similar values of the saturation magnetization $M_{s}$ for the different field directions confirm rather isotropic $g$-factors: 7.1(1) ($B\|ab$) and 7.0(1)~\mbfu\ ($B\|c$). The saturation fields amount to $B_{\rm s}^{ab} = 7.3(2)$~T and $B_{\rm s}^{c} = 8.4(2)$~T. We also note several features in the $M(B)$ curves, the most prominent one appearing at around $M_s$/3 for \bpc~$\simeq 2.9$~T.

At $T= 0.4$~K, the anomalies in the magnetization curves are most pronounced as displayed in Fig.~\ref{M400mK} (full $M(B)$ curves covering -7~T~$\leq B\leq$~7~T are shown as Fig.~S2 in the SM~\cite{SM}). The main features are as following: (1) There is a small ferromagnetic moment as well as magnetic hysteresis visible in the inset of Fig.~\ref{M400mK}a which show, for $B||c$, a small remanent moment of 0.14~\mbfu\ and the critical field $\simeq 60$~mT. No indication of hysteresis is found for $B||ab$ which however displays s-shaped behaviour around $B=0$~T as shown in the inset in Fig.~\ref{M400mK}b and by the broad peak in $\partial M/\partial B_{||ab}$ centered at $B=0$~T. (2) For \bpc , there is a plateau-like feature in $M$ centered at \bcone~= 2.9~T and a small jump in $M$ at \bctwo~= 5.5~T. Above \bctwo , $M(B)$ features linear behavior which extrapolates to its saturation value at \bs~$\simeq 8.6$~T~\footnote{Note, that applying the same method at 1.8~K where the saturation field has been determined experimentally (see Fig.~\ref{Mintro}), i.e., extrapolating the linear regime at 1.8~K, yields a value of \bs\ which agrees well with the saturation field derived from the actual measurements up to 14~T at this temperature.}. We also note a feature at \bconeupper~= 4.2~T signaling the onset of the linear-in-$M$ regime and thus the upper limit of the plateau region while the plateau's lower edge is marked \bconelower . (3) For \bpab , there is a jump in $\partial M/\partial B$ at \baone\ $\simeq 2.7$~T which is preceeded by a tiny peak (i.e., a small jump in $M$) at \baone\ and followed by a linear regime in $M$ for $B>B_{\rm a2}$.

To summarize the main features, there is a clear magnetization plateau visible in $M(B||c)$, extending from \bconelower\ to \bconeupper\ and centered at \bcone , which perfectly agrees to 1/3 of the saturation magnetization as determined at 2~K (cf. Fig.~\ref{Mintro}d). Note, that we find $M_{\rm s}$ rather independent on temperature in the accessible temperature regime so that $M_{\rm s}$(2~K)~$\simeq~$$M_{\rm s}$(0.4~K). A similar conclusion on the presence of a 1/3 magnetization plateau has been drawn from magnetization data at 1.8~K in Ref.~\cite{GIO2021magnetic}. In addition, there is a small jump in $M$ at \bctwo\ signaling a discontinuous phase transition at about $3/5M_{\rm s}$. We also note that the tiny peak in $\partial M/\partial B_{||ab}$ at \baone\ appears at 2/5 of the saturation magnetization and \batwo\ which signals the onset of a linear regime in $M(B>$\batwo$)$ appears at $\simeq$1/2 of $M_s$.

\subsection{Magnetic phase diagrams\label{section:phd}}

\begin{figure}[htb]
\centering
\includegraphics [width=\columnwidth,clip] {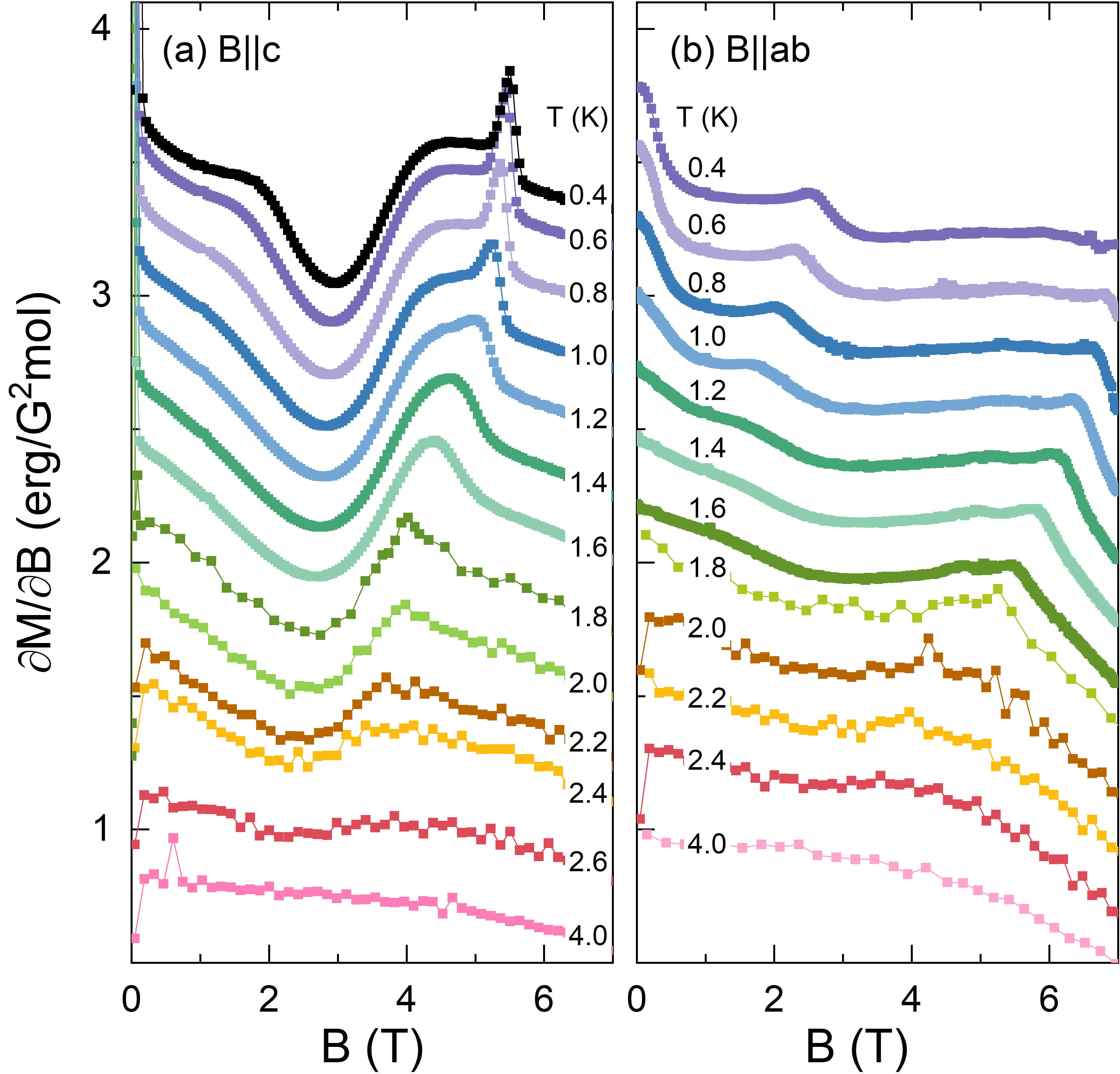}
\caption{Magnetic susceptibility $\partial M/\partial B$ for (a) \bpc\ and (b) \bpab\ at different temperatures. The curves are offset vertically by \E\ for better visibility.} \label{fig.6}
\end{figure}

\begin{figure*}
\centering
\includegraphics [width=2\columnwidth,clip] {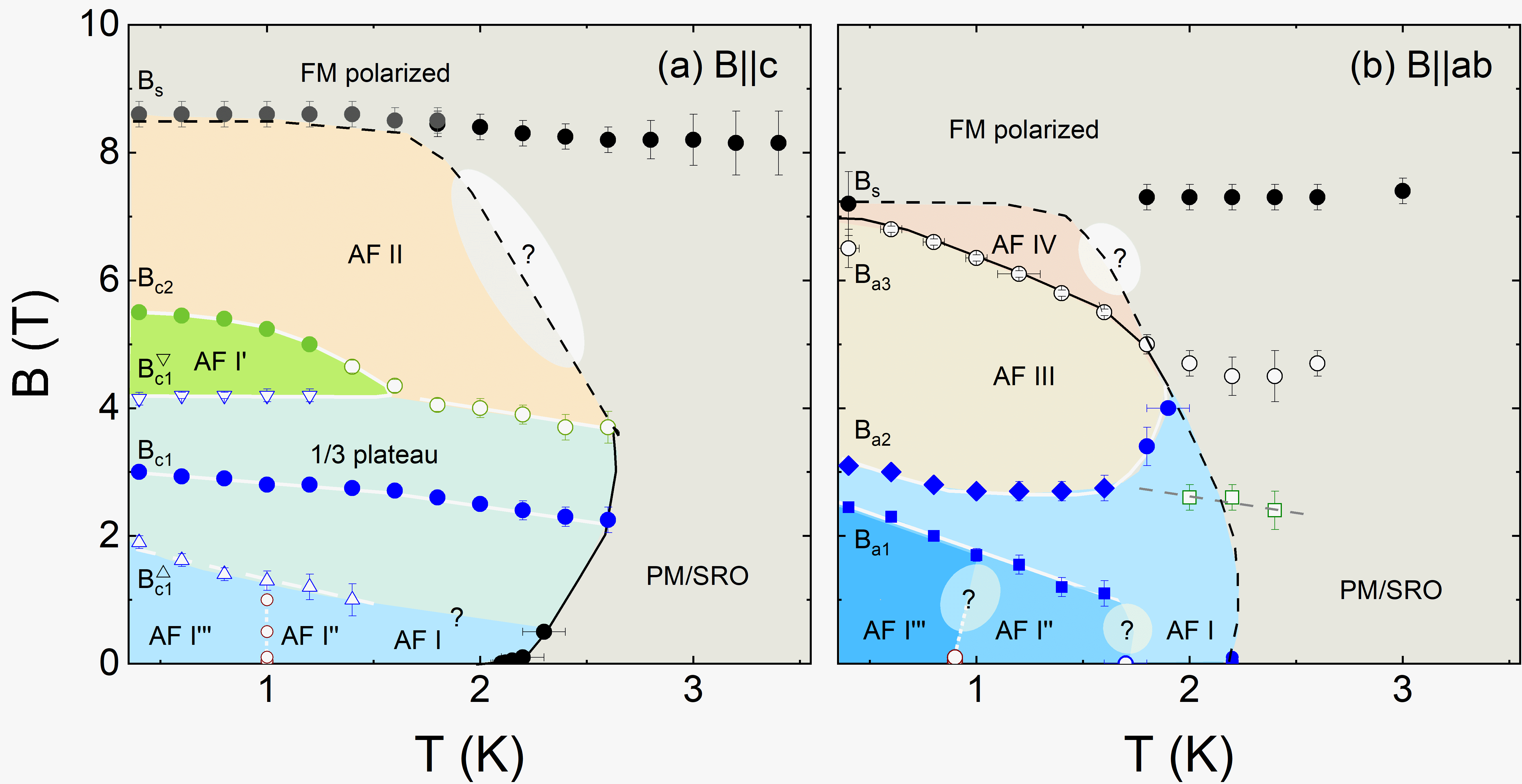}
\caption{Magnetic phase diagram of \gio\ for $B||c$ and $B||ab$. $B_{{\rm a/c}i}$ mark the critical fields as defined in Fig.~\ref{M400mK} and in the text. $B||c$: AF I: Low-field AF phase; 1/3 plateau phase centered around \bcone\ and confined by \bconelower\ and \bconeupper . The latter marks the onset of a linear-in-$M$ regime AF I${'}$. At \bctwo\ a small jump in $M$ appears before the polarized regime is achieved at \bs . $B||ab$: At $B=0$~T, several features distinguish regions of different spin-orientation (AF I, AF I${''}$, AF I${'''}$). A clear phase boundary \baone ($T$) separates AF I${'''}$ from the higher-field phase which may extend to AF I. Two AF phases (AF I/AF III) are separated by \batwo . A further kink in $M$ marks the onset of a high-field but not fully polarised phase AF IV at \bathree . Several features appearing in the paramagnetic/short-range ordered phase (PM/SRO) are shown, too, as well as features the regimes AF I, AF I${''}$, and AF I${'''}$. White areas mark regions where the phase boundaries are yet unclear.} \label{phd-rk}
\end{figure*}

Distinct anomalies in $M$ vs. $B$ allow us to trace the temperature dependence of the phase boundaries associated with the critical fields marked in Fig.~\ref{M400mK}. Specifically, we have used the anomalies of the magnetic susceptibility shown in Fig.~\ref{fig.6} to construct the magnetic phase diagrams. In addition, we have derived the saturation fields from our $M(B,T\geq 1.8~\rm{K})$ data up to 14~T (see Fig.~\ref{Mintro}d) and from extrapolating the linear-in-field behaviour at 0.4~K~$\geq T\geq$~1.8~K, as well as \tn ($B$), $T^{*}(B)$, and $T^{**}(B)$ at low fields from $M$ vs. $T$ measurements (see Fig.~\ref{chicp}). The resulting phase diagrams are shown in Fig.~\ref{phd-rk}.

The following main features appear for $B||c$: Centered at \bcone , a 1/3 magnetization plateau is formed which starts to evolve at \bconelower\ and extends to \bconeupper . \bcone\ is barely temperature dependent as it only slightly shifts to lower fields upon heating. Below about 1.2~K, there is a linear-in-$B$ regime of the magnetization following the AF~I/plateau phase. Whether the AF I/AF I${'}$/AF I${''}$ boundaries signal proper thermodynamic phase transitions or crossover regimes is yet unclear. A sharp jump in $M$ at \bctwo\ clearly indicates a discontinuous phase transition and suggests a flip of the spin configuration. This jump is superimposed by a kink in $M$ vs. $T$ as demonstrated by the peak and superimposed jump in $\partial M/\partial B$ in Fig.~\ref{M400mK}. Upon heating, the sharp peak in $\partial M/\partial B$ evolves to a broader feature in the temperature region where the upper boundary of the plateau phase (\bconeupper ) merges with \bctwo . Broadening and softening of the anomaly may indicate that the phase boundary evolves towards a continuous nature which would suggest the presence of a tricritical point at $\simeq 4$~T and $\simeq 1.6$~K.

Clausius-Clapeyron equation enables us to estimate the entropy changes appearing at the AF I$'$/AF II phase boundary~\cite{Barron}:
\begin{equation}
    \Delta S_{\rm c2}=- \Delta M_{\rm c2}\times \frac{\partial B_{\rm c2}}{\partial T}. \label{clausius}
\end{equation}
Using $\Delta M(\rm{0.4~K})\simeq 0.11(1)$~\mbfu , the analysis yields $\Delta S_{\rm c2}\simeq 0.15(3)$~\jmk , at 0.4~K.

As described above, for $T>1.2$~K, the sharp peak transforms into a much broader feature and \bctwo\ is suppressed upon further heating. The saturation field towards the ferromagnetically polarized phase does not display strong temperature dependence and the saturation features can be traced well above the long-range ordered phase. We attribute this to significant short-range magnetic correlations above \tn . In contrast to \bs , magnetization measurements do not detect the phase boundary \tn ($B>4$~T) indicating that the magnetization in AF II and in the short-range ordered phase is very similar in this field regime. One may speculate about the phase boundary as suggested in Fig.~\ref{phd-rk}a.

We note that the magnetization $M_c$ increases at \tn ($\simeq$0~T) which already implies the observed positive initial field dependence of the associated phase boundary. Quantitatively, it may be estimated using the Ehrenfest equation:
\begin{equation}
    \frac{\partial T_{\rm N}}{\partial B}=-T_{\rm N}\frac{\Delta M'}{\Delta c_{\rm p}}.\label{ehrenfest}
\end{equation}
Exploiting the experimentally determined jump in specific heat $\Delta c_{\rm p}$ from Fig.~\ref{chicp}a and the change in slope of magnetization $\Delta M'=\Delta (\partial M/\partial T)$ yields $\partial T_{\rm N}/\partial  B_{||c} = 1.8(2)$~K/T which is consistent with our data \tn ($B\gtrsim 0$~T) shown in Fig.~\ref{phd-rk}a.

For $B||ab$ (Fig.~\ref{phd-rk}b), we observe no sizable field dependence of \tn\ in small magnetic fields. This agrees to the observation of only a small increase of $M_{ab}$ at \tn . The quantitative analysis in terms of the Ehrenfest relation (Eq.~\ref{ehrenfest}) yields only $\partial T_{\rm N}/\partial B~\simeq 0.02(1)$~K/T. At zero field, two further anomalies in $M(T)$ (Fig.~\ref{chicp}b) indicate rotation of the ferromagnetic component. The related regimes in the phase diagram are labelled AF I$''$ and AF I$'''$ and the nature of boundaries is not clear (Fig~\ref{phd-rk}b).

 A small peak and a subsequent jump in $\partial M/\partial B_{||ab}$ (see Fig.~\ref{M400mK}b) signals the appearance of a high-field phase AF~III. From the fact that both anomalies \baone\ and \batwo\ further separate upon heating, we conclude that the intermediate phase extends to the AF I regime at $B=0$~T. While AF I is separated by distinct anomalies from AF III around 2~K, \baone\ cannot be traced up to \tn . \tn ($B$) cannot be well traced by our data either. In contrast, the upper phase boundary of AF~III, i.e., \bathree , is marked by clear anomalies in $M(B)$. From the fact that \bs\ seems to be rather independent on temperature and clearly exceeds/is distinct from \bathree , we conclude the presence of the high-field phase AF IV~\footnote{Extrapolating the experimental data $M(B\leq 7~{\rm T})$ to higher fields as done for $B||c$ in order to estimate \bs\ only yields ambiguous results due to pronounced bending of $M(B)$ in this regime. We hence do not apply this method for $B||ab$.}. Note, that again several features in $M(B)$ extend into the short-range ordered/paramagnetic phase.

\begin{figure*}[htb]
\centering
\includegraphics [width=2\columnwidth,clip] {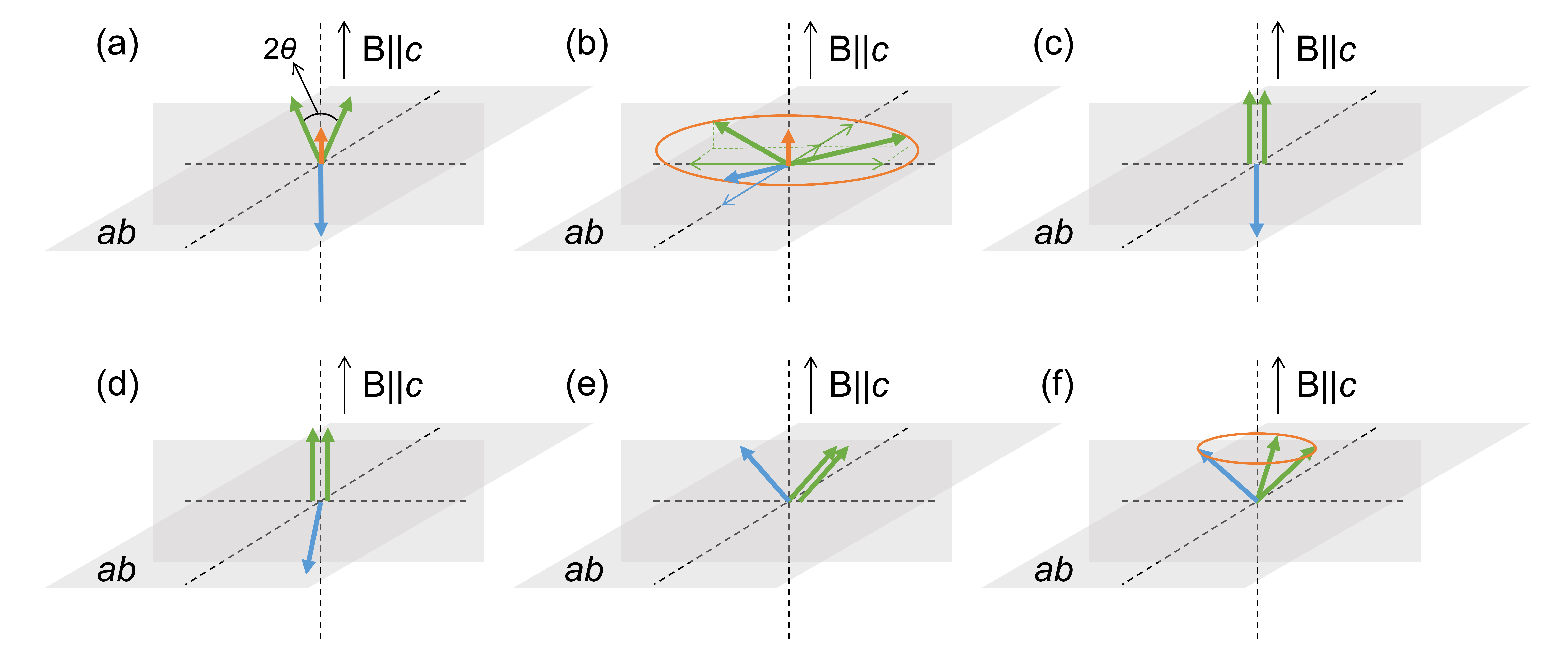}
\caption{Schematic diagram of possible spin configurations in \gio\ in different magnetic fields. The green and blue arrows indicate the spin directions at the Gd1 and Gd2 positions, respectively, which correspond to the Gd atom positions in Fig.~\ref{fig.3}. The orange arrows indicate the net magnetic moment along $c$ axis. The angle between the upward spins in (a) is denoted by $2\theta$.} \label{sketch}
\end{figure*}

Finally, we discuss the phase diagram in the frame of potential spin configurations appearing in triangular-lattice spin systems (see, e.g., \cite{CCB2003plateau,RFMO2006plateau,BCSO2013plateau,Starykh2015review}). The ground-state spin configuration has not yet been determined experimentally. Potential candidates of the ground state configuration are coplanar Y-type (as also discussed in Ref.~\onlinecite{GIO2021magnetic}) or umbrella-type configurations. In both cases, finite-temperature $uud$ phases may appear~\cite{Griset2011,Starykh2015review}.

(1) Our experimental data imply the presence of a net magnetic moment. In case of the Y-type configuration, the measured net moment would correspond to an angle of $2\theta \simeq 117^\circ$ between the upper spins in Fig.~\ref{sketch}a. An umbrella structure would feature a huge aperture outside the $ab$ plane (see Fig.~\ref{sketch}b); the observed size of the canted moment would suggest an angle of $\simeq 1.2^{\circ}$ between the Gd moments and the $ab$ plane. The in-plane projections of the moments cancel out in this scenario. The data in Fig.~\ref{chicp} indicate partial rotation of the net moment towards the $ab$ plane, yielding further distortions of the above mentioned configurations. From Fig.~\ref{chicp}c we conclude that this rotation appears in two steps at $\simeq 1.7$~K and $\simeq 1$~K, respectively. Small external magnetic field $B||c$ yields a ferromagnetic-like hysteresis (see Fig.~\ref{M400mK}a) with the $c$ axis being the magnetic easy axis of the net moment.

(2) Applying intermediate fields $B||c$ yields the formation of a colinear $uud$ configuration (Figs.~\ref{sketch}c), i.e., the plateau phase. 
At 0.4~K, the plateau region is centered at 2.9~T and extends from 1.9 - 4.2~T (Fig.~\ref{M400mK}a). Our finding of a smeared-out and not completely flat plateau only for $B||c$ does support the scenario of a classical Heisenberg triangular-lattice antiferromagnet (TLAF) with easy axis anisotropy~\cite{miyashita1986japan} as discussed, e.g., for Na\(_2\)BaCo(PO\(_4\))\(_2\)~\cite{NBCPO2020plateau}, Rb\(_4\)Mn(MoO\(_4\))\(_3\)~\cite{RMMO2011plateau}, GdPd\(_2\)Al\(_3\)~\cite{GPA1999}, Ba\(_3\)MnNb\(_2\)O\(_9\)~\cite{BMNO2014plateau} and Ba\(_3\)NiSb\(_2\)O\(_9\)~\cite{BNSO2011quantum}. The center and edges of the plateau phase do not strongly change upon heating. However, the plateau significantly blurs and no clear signature of \bconelower\ can be identified above $\simeq 1.5$~K.

(3) Above \bconeupper , the linear increase in $M$ implies breaking of the $uud$ configuration and the continuous alignment of spins towards the field (Fig.~\ref{sketch}d). A similar behaviour is predicted in Ref.~\cite{miyashita1986japan}; in agreement with these numerical studies, phase AF I$'$ is found to be destabilised in external fields $B||c$.

(4) In contrast to the predictions of the minimal TAF model~\cite{miyashita1986japan}, we observe an additional discontinuity at \bctwo\ associated with a jump-like increase of magnetization. The jump is from about 4.3 to nearly 4.5~\mbfu , i.e., it starts at about 3/5 of the full saturation magnetization. In this magnetization regime, there are several possible scenarios which may account for such an behaviour. One of which includes discontinuous rotation of the $uu$-moments from the easy direction towards a coplanar V-shaped structure which may evolve from the $uud$ phase by decreasing the angle $\angle(uu,d)$ as sketched in Fig.~\ref{sketch}e. However, our data do not allow to unambiguously resolve the spin configurations in this field range.

(5) For \bpab , no plateau phase is formed. Instead, we observe a small kink and a jump in $\partial M/\partial B$, at \baone\ and \batwo\ (see Fig.~\ref{M400mK}b). Note, that the magnetization at \baone\ and \batwo\ amounts to 2.85(5)~\mbfu\ and 3.50(5)~\mbfu , respectively, which is very similar to 2/5 and 1/2 of the saturation values. Due to the presence of several rotated spin arrangements (rotations at $\simeq 1.7$~K and $\simeq 1$~K), only rough speculations on the field effects are possible. The behaviour of $M(B||ab)$ around $B=0$~T however shows that the net magnetic moment is rather smoothly aligned into the $ab$ plane. The origin of the further distinct phases in the magnetic phase diagram Fig.~\ref{phd-rk}b remains to be clarified.


In TLAFs with classical spins, both easy-axis anisotropy and easy-plane anisotropy can stabilize the $uud$ phase at finite temperatures (see, e.g., \cite{BCNO2014plateau,BMNO2014plateau}). The presence of the 1/3 plateau in \gio\ implies that such easy axis/easy plane anisotropy which lifts geometric frustration is relevant for driving the system into the $uud$ configuration. Comparing the temperature evolution of the $uud$ phase for $B||c$ with numerical studies~\cite{miyashita1986japan} suggests that \gio\ has a weak easy-axis anisotropy. This conclusion is corroborated by the absence of the 1/3 magnetization plateau for \bpab . It is also in-line with model calculations of the magnetization of \gio\ in Ref.~\cite{GIO2021magnetic} where weak easy-axis anisotropy $D>0$ and $J_{\rm 1}$ $\geq$ $J_{\rm 2}$ are suggested.

Significant anisotropy might arise from dipolar interactions between the Gd$^{3+}$ moments within the filled honeycomb layer. Dipole-dipole interaction can be described by the Hamiltonian~\cite{bencini2012epr}:
\begin{equation}
	\mathbf{\mathcal{H}}_\mathrm{dip}=g^2\mu_\mathrm{B}^2\sum_{i<j} ({\mathbf{S_i}\mathbf{S_j}/r^3_\mathrm{ij}-3(\mathbf{S_i}\mathbf{r_{ij}})(\mathbf{S_j}\mathbf{r_{ji}})/r_\mathrm{ij}^5 )}
\end{equation}
with $\mathbf{r_{ij}}$ being the vector between interacting paramagnetic centers and $r_\mathrm{ij}$ the distance between them, such that the dipolar anisotropy $D_\mathrm{dip}$ can be estimated as $D_{dip}=E_{dip}/S^2=\mu_\mathrm{B}^2p_\mathrm{eff}^2\mu_0/4\pi r^3S^2$. The resulting dipole-dipole energy amounts to $E_\mathrm{dip}=\mu_\mathrm{B}^2p_\mathrm{eff}^2\mu_0/4\pi r^3=0.841(2)\,\mathrm{K}$, with the weighted average $r_\mathrm{ave}=(2r_1+r_2)/3=3.6753(5)$ calculated from the distances $r_1=3.6813(5)$ and $r_2=3.6633(5)$ from our structure refinement. This energy corresponds to $T^{*}$ and $T^{**}$ where reorientation processes are observed. Similar anisotropy energies for Gd$^{3+}$-systems were reported in the literature  for (Y$_{1-x}$Gd$_x$)$_2$Ti$_2$O$_7$~\cite{GTO2005single} and Gd$_2$Ti$_2$O$_7$~\cite{GTO1999transition}.

\section{Summary}

By means of single crystals grown by the high-pressure optical floating-zone method, the magnetization process down to the mK-regime and the magnetic phase diagrams of \gio\ have been investigated. The system evolves long-range antiferromagnetic order at \tn\ = 2.1~K and exhibits considerable magnetic frustration ($\lvert\Theta\rvert /T_{\rm N}\simeq 5$). The ground state features a small net magnetic moment  along the crystallographic $c$ direction which reorients upon cooling at $T^{**}\simeq 1.7$~K and $T^{*}\simeq 1$~K. A broad 1/3 plateau indicative of the $uud$ spin configuration appears for $B||c$ but is absent for $B||ab$, thereby suggesting easy axis anisotropy. In this respect \gio\ is a typical triangular lattice material in which weak easy-axis anisotropy of presumingly dipolar nature breaks $C_{\rm 3}$ symmetry and allows formation of the $uud$ phase. In addition, a jump in magnetization at low temperatures signals a discontinuous transition to a high field phase. There is evidence that the transition evolves a continuous nature upon heating via a possible tricritical point. Small energy and field scales in the accessible regimes render \gio\ a prime example to study the phase diagram of a semiclassical frustrated hexagonal lattice in the presence of weak easy axis anisotropy, e.g., by future neutron diffraction studies to verify the actual spin configurations in the various thermodynamic phases.

\begin{acknowledgments}

Support by Deutsche  Forschungsgemeinschaft (DFG) under Germany’s Excellence Strategy EXC2181/1-390900948 (The Heidelberg STRUCTURES Excellence Cluster) and through project KL 1824/13-1 is gratefully acknowledged. N.Y. acknowledges fellowship by the Chinese Scholarship Council (File No. 201906890005).
\end{acknowledgments}

\bibliography{GIO3-1}

\end{document}